\documentstyle[prb,preprint,aps,psfig]{revtex}

\begin{document}

\draft \title{Cohesive energies  of  cubic III-V semiconductors}

\author{Beate Paulus, Peter Fulde} \address{Max-Planck-Institut f\"ur
Physik komplexer Systeme, Bayreuther Str. 40, 01187 Dresden, Germany}
\author{Hermann Stoll} \address{Institut f\"ur Theoretische Chemie,
Universit\"at Stuttgart, 70550 Stuttgart, Germany}

\maketitle

\begin{abstract} Cohesive energies for twelve cubic III-V
semiconductors with zincblende structure have been determined using
an ab-initio scheme.  Correlation contributions, in particular, have
been evaluated using the coupled-cluster approach with single and
double excitations (CCSD). This was done by means of increments obtained for
localized bond orbitals and for pairs and triples of such bonds.
Combining these results with corresponding Hartree-Fock  data, we
recover about 92 \% of the experimental cohesive energies.
\end{abstract} 

\section{Introduction}

{\em Ab-initio} calculations for ground state properties of solids
are most often performed nowadays within the density-functional (DFT)
scheme.  Typically the exchange and correlation contributions are
calculated within the local-density approximation (LDA)
\cite{jones89}, or, more recently, within the generalized-gradient
approximation (GGA) \cite{perdew95}. These methods yield quite good
results for cohesive energies and other properties, but do not
provide many-body wave functions; thus, a systematic improvement
towards the exact result appears to be difficult.\\ Hartree-Fock
self-consistent-field (HF-SCF) calculations for solids, on the other
hand, while lacking electron correlation as per its definition, have the
merit of treating non-local exchange exactly and of supplying a good
starting point for subsequent correlation calculations. One
possibility suggested in literature is to calculate correlation
contributions on top of HF within LDA or GGA \cite{causa94}, but the
potential advantage through the availability of a many-body
wavefunction is absent and the results do not seem to achieve a
significant improvement over exchange-correlation(xc)-DFT. \\ For
finite systems like atoms and molecules, accurate quantum-chemical
methods such as configuration-interaction (CI) or coupled-cluster
(CC) approaches have been developed for the determination of
correlation energies.  Infinite systems such as solids require the
use of size consistent approximations for treating electron
correlations like coupled electron pair approximations (CEPA) or CC
approaches. Because of the local character of the correlation hole
one can expand the correlation energy in terms of local increments
\cite{stoll92}.  The idea is thereby to determine the required matrix
elements by studying local exitations in clusters which are
accessible to a full quantum-chemical treatment. A formal derivation
of the method for an infinite periodic system can be found in Ref. [5] or
[6] when the CEPA scheme is used.  The derivation for the CC method
is very similar, see, e.g., Ref. [7].  The method of increments has
previously been applied to ground-state properties (cohesive
energies, lattice constants and bulk moduli) of elementary
semiconductors and has led to very good and reliable results
\cite{paulus95}.\\ In the present paper we extend the method to polar
semiconductors, especially to the twelve cubic III-V compounds. In
order to assess the validity of the calculated correlation energies,
we evaluate their differential contributions to the cohesive energies
of these materials. For a comparison with experiment, however, we
also need reliable SCF ground state energies.  In Section II we
perform the SCF calculations using the program package {\sc
Crystal92} \cite{crystal92}. In Section III we sketch the method of
correlation-energy increments and report on computational details.
The results are presented and discussed in Section IV. Conclusions
follow in Section V.

\section{Hartree-Fock Calculations}

For discussing the influence of electron correlation effects in
solids, reliable SCF calculations are prerequisites. For the
compounds which we are interested in, SCF calculations with the program
package {\sc Crystal88} have been performed by the Torino
group\cite{causa91}. However small split-valence basis sets have
been used in these calculations, and it is not clear how closely the
results approach the HF limit.\\ Here, we want to supplement their
results with our own calculations employing the scalar-relativistic
energy-consistent pseudopotentials (PP) of the Stuttgart group
together with the corresponding atomic basis sets\cite{bergner93}.
The cores replaced by the pseudopotentials are X$^{3+}$ and X$^{5+}$
for group 13 and 15, respectively, with two exceptions:
3-valence-electron pseudopotentials for the post-$d$ elements Gallium
and Indium have been found to underestimate the closed-shell
repulsion of the underlying $d$ shell on valence electrons of
neighbouring atoms\cite{schwerdt95,leining95}; we therefore performed
the SCF calculations for Ga and In compounds with 13-valence electron
pseudopotentials\cite{13vepp}, explicitly treating the highest
occupied $d$ shell.\\ For the solids, we generated
$(4s4p1d)/[3s3p1d]$ Gaussian valence basis sets (Ga: from the atomic
basis set\cite{13vepp} $(8s6p7d)$ the very diffuse exponents are
neglected in obtaining $(7s4p6d)/[3s3p2d]$, In:  from the atomic basis
set\cite{13vepp} $(6s6p6d)$ the very diffuse exponents are neglected
in obtaining $(4s4p5d)/[3s3p2d]$) as follows: starting from the
energy-optimized atomic basis sets derived for the PP just
mentioned\cite{bergner93,13vepp}, the inner two (Ga: five) functions
of $s$ and $p$ symmetry were contracted using atomic ground-state
orbital coefficients.  The outer two $s$ and $p$ functions were left
uncontracted.  The most diffuse exponents of the atom have little
effect in the solid, where, due to close-packing, their role is
taken over by  basis functions of
the neighboring atoms.  Moreover, the use of 
basis functions that are too diffuse leads to numerical problems in {\sc
Crystal}.  Therefore we re-optimized the two outer $s$ and $p$
functions as well as an additional $d$ polarization function for the
solid.  The crystal-optimized basis sets are listed in Table
\ref{basis}.\\ Using the above basis sets, we determined SCF ground
state energies of the solid with {\sc Crystal92} \cite{crystal92}.
When evaluating cohesive energies, we subtracted corresponding
energies of free ground-state atoms obtained with the original
atom-optimized sets\cite{bergner93,13vepp}, using the
quantum-chemical {\em ab-initio} program {\sc
Molpro94}\cite{molpro94}.  The results for the SCF cohesive energies
at experimentally determined values of lattice constants 
\cite{landoltiii17a} are listed
in Table \ref{result}, and comparison is made to experimental
cohesive energies corrected by the phonon zero-point energies
$\frac{9}{8}{\rm k_B}\Theta_{\rm D}$ (derived from the Debye
model\cite{farid91}) as well as by atomic spin-orbit
splittings\cite{moore}.\\ It is seen that binding energies, at the HF
level, are between 50\% and 70\% of the experimental values leaving
room for significant correlation contributions.  The agreement with
the previously published SCF cohesive energies\cite{causa91} is very
good for compounds containing B and Al, while small deviations of
$\sim$2\% of the cohesive energy are found for Ga and In compounds.
The stability of the results with quite different pseudopotentials
and basis sets (Causa et al used PP's of the Toulouse group and
[$2s2p1d$] basis sets\cite{causa91}) is gratifying to note, and we
expect to be not very far from the basis-set limit of the SCF
cohesive energy with our [$3s3p1d$] valence basis sets.

\section{Correlation effects}

Calculating correlation effects using an expansion in terms of
local increments\cite{stoll92} is formally similar to treating the
hierarchy of $n$th order atomic Bethe-Goldstone equations\cite{BG}.
Here we only want to sketch the basic ideas and some important
formulae (for more details see Ref. [5] and [6]).  The method relies
on localized bond orbitals generated in a SCF reference calculation.
One-bond correlation-energy increments $\epsilon_i$ are obtained by
correlating each of the localized orbitals separately while keeping
the other ones inactive. In the present work we are using the
coupled-cluster approach with single and double substitutions
(CCSD).  This yields a first approximation of the correlation energy
\begin{equation} E_{\mbox{corr}}^{(1)}=\sum_i \epsilon_i ,
\end{equation} which corresponds to the correlation energy of
independent bonds.\\ In the next step we include the correlations of
pairs of bonds. Only the non-additive part $\Delta\epsilon_{ij}$ of
the two-bond correlation energy $\epsilon_{ij}$ is needed.
\begin{equation}
\Delta\epsilon_{ij}=\epsilon_{ij}-(\epsilon_i+\epsilon_j).
\end{equation} Higher order increments are defined analogously. For
the three-bond increment, for example, one has \begin{equation} \Delta
\epsilon_{ijk}= \epsilon_{ijk} -( \epsilon_i +  \epsilon_j +
\epsilon_k) -(\Delta \epsilon_{ij}+\Delta \epsilon_{jk}+ \Delta
\epsilon_{ik}).  \end{equation} The correlation energy of the solid
is finally obtained by adding up all the increments with appropriate
weight factors:  \begin{equation} E_{\mbox{corr}}^{\mbox{solid}} =
\sum_i  \epsilon_i + \frac{1}{2}\sum_{ij \atop i \not= j} \Delta
\epsilon_{ij}+ \frac{1}{6}\sum_{ijk \atop i \not= j \not= k} \Delta
\epsilon_{ijk} + ... .  \end{equation} It is obvious that by
calculating higher and higher increments the exact correlation energy
within CCSD is determined.\\ The procedure described above is only
useful if the incremental expansion is well convergent, i.e. if
increments up to, say, three-bond increments are sufficient, and if increments
decrease rapidly with increasing distance between localized
orbitals.  These conditions were shown to be well met in the case of
elementary semiconductors \cite{paulus95}, but have to be checked
again for the polar compounds.\\ Since (dynamical) correlation is a
local effect, the increments should be fairly local entities at least
for semiconductors and insulators.  We use this property to calculate
the correlation-energy increments in finite clusters.  We select the
clusters as fragments of the zincblende structure so that we can
calculate all two-bond increments up to third nearest neighbors and
all nearest-neighbor three-bond increments.  The clusters are shown
in Fig.\ref{fig}. The bond length between the group 13 (X) and the
group 15 atom (Y) is taken to be the same as in the solid at the
experimental equilibrium lattice constant. The dangling bonds are
saturated with hydrogens. The X---H and the Y---H distances,
respectively, were optimized in CCSD calculations for ${\rm XYH}_6$
clusters, yielding $d_{\rm BH}$= 1.187 \AA , $d_{\rm NH}$= 1.016 \AA
, $d_{\rm AlH}$= 1.614 \AA , $d_{\rm PH}$= 1.408 \AA , $d_{\rm GaH}$=
1.621 \AA , $d_{\rm AsH}$= 1.525 \AA ; for In and Sb, we adopted the
SnH$_4$ distance (1.711 \AA ).  Instead of constructing localized
bond functions from the Bloch states derived using the {\sc Crystal92}
program, it is simpler to determine them from a cluster
calculation. One merely has to show that, for a given increment,
convergency is reached when the size of the cluster is sufficiently
large.  With this in mind we performed standard SCF calculations for
each cluster and localized the bonds according to the Foster-Boys
criterion \cite{foster60} within the occupied valence space in ${\rm
C}_1$ symmetry. Following the procedure described above we calculated
the correlation-energy increments at the CCSD level (using the
program package {\sc Molpro94}\cite{molpro94}) successively
correlating more and more of the localized X---Y bonds.\\ For
hydrogen we chose Dunning's \cite{dunning89} double-zeta basis
without a polarization function.  Two different basis sets are used for
the other elements.  Basis A is of the same quality as for the SCF
calculations of the preceding section:  $(4s4p)/[3s3p]$
\cite{bergner93} valence basis sets energy-optimized for the atoms,
with $d$ polarization functions optimized in CCSD calculations for
${\rm XYH}_6$ clusters (see Table \ref{polbasis}).  Extended basis
sets (B) have been generated by uncontracting the $sp$ functions of
basis A and by replacing the single $d$-function  by a $2d1f$
polarization set.  The latter was obtained by adding
 a second more diffuse $d$ function
to the one of basis A and by optimizing the $f$ exponent in CCSD
calculations for the ${\rm XYH}_6$ clusters (see Table
\ref{polbasis}).\\ We have checked the convergence of the incremental
expansion for all polar substances and will discuss it here for GaAs.
As can be seen from Table \ref{inc}, the increments decrease quite
rapidly with increasing distance of the bonds. The nearest neighbor
two-bond increments contribute 51\% of the correlation energy of GaAs, 
all next-nearest neighbors 13\% and the third-nearest
ones only 6\% . After these increments we truncate the expansion because
the fourth-nearest neighbors' contribution is only 0.5\% . Note that
the integrated truncation error decreases with $r^{-3}$,
because the interaction is van der Waals like $r^{-6}$
and the number of pairs of bonds grows like $r^2$.\\ The
convergence with respect to the number of bonds correlated
simultaneously in the incremental expansion is also quite
satisfactory. For GaAs again, the one-bond increment contributes
38\% of the correlation energy, the two-bond 
increments are very important (70\%) but yield a
correlation energy too large in magnitude, while the three-bond increments
reduce it by 8\%.  An estimate of the nearest-neighbor four-bond
increments and next-nearest neighbor three-bond increments is
$\sim$0.5\% .  Thus, the overall error due to truncations in the
incremental expansion is less than 1\% of the correlation energy, the
same accuracy as was reached for the elementary
semiconductors\cite{paulus95}.\\ The next point we have to check is
the transferability of the increments from clusters to the solid.  We
test the sensitivity of the chemical surroundings by determining the
increments in different clusters.  The change from ${\rm GaAsH}_6$,
where the Ga---As bond is surrounded only by hydrogens, to ${\rm
Ga}_4{\rm As}_4{\rm H}_{18}$, where the inner Ga---As bond has the
same neighbors as in the solid, is 0.005 a.u.\  for the one-bond
increment.  The differences in the nearest-neighbor  two-bond
increments are -0.005 a.u., so that the main errors due to lack of
transferability cancel. The tendency for individual (one- and
two-bond) errors to be of different sign is seen in all III-V
compounds. We estimate an upper limit for the transferability error
to 1\% of the correlation energy.\\ Thus, we end up with a total
error of 2\% for approximations inherent in our use of the
incremental expansion of the correlation energy.  Not included are
shortcomings of one- and many-particle basis sets in the
determination of individual increments; the largest part of this
remaining error is probably due to limitations of the one-particle
basis set and will be discussed in the next section.

\section{Results and discussion}

Applying the method of increments as described in the preceding
section, we determined correlation contributions to cohesive energies
for all twelve cubic III-V semiconductors. The increments were always
taken from the largest possible cluster (cf.\ Table III and Fig.
\ref{fig}) and multiplied by the weight factors appropriate for the
zincblende structure.  The correlation contributions to the cohesive
energies were obtained as $ E_{\rm coh}^{\rm corr}=E_{\rm solid}^{\rm
corr}- \sum_i E_{{\rm atom},i} ^{\rm corr}$ per unit cell.  The
results for the two different basis sets are shown in Table
\ref{result}. For basis set A we obtain an average of 86\% of the
experimental cohesive energies, which amounts to $\approx$67\% of the
'experimental' correlation contributions to the cohesive energies
(defined here as the differences between the experimental cohesive
energies and the corresponding SCF values).  The larger basis set B
yields  a  substantial improvement, $\approx$82 \% of the correlation
contributions to the cohesive energies.  Overall we obtain with basis
B an average of 92\% of the cohesive energy with only small
fluctuations for the different materials. In earlier calculations for
the elementary semiconductors\cite{paulus95}, we obtained $\approx$94\%
of the cohesive energy.  The deviation may be due, in part, to the
different correlation treatment. For the elementary semiconductors we
used a coupled-electron-pair approximation (CEPA-0).  It is well
known that CEPA-0 yields larger correlation contributions as the CCSD
approach; in test calculations we found a deviation of $\approx$1\%.  If
the basis set, especially the polarisation set is enlarged even more,
e.g., to $3d2f1g$, a further increase of the correlation effects can
be expected.  In test calculations for GaAs, where the one-bond
increment and the nearest neighbor two-bond increments were
determined with this extended basis set, we obtained an increase of
10\% of the correlation contribution to the cohesive energy.\\ For
comparison, we have also listed in Table \ref{result} results from
the literature which have been obtained with other methods.  LDA
\cite{causa94} overestimates the cohesive energies by $\approx$20 \% ,
GGA is at the same level of accuracy or slightly better than our
results. In contrast to these methods, our approach lends itself to
systematic improvement; moreover, a detailed insight into the nature
of the correlation effects becomes possible.

\section{Conclusions}

We have determined cohesive energies of twelve cubic III-V
semiconductors, both at the SCF and the CCSD levels.  The SCF results
have been obtained with the {\sc Crystal92} code using relativistic
energy-consistent pseudopotentials and  $[3s3p1d$] Gaussian valence
basis sets.  Electron correlations are described at the CCSD level
with the method of local increments, using the same pseudopotentials
in cluster calculations with the {\em ab-initio} program {\sc
Molpro94}. Increasing cluster and basis-set size, this approach
allows for a systematic improvement of accuracy towards the fully
correlated solid-state limit.  The results show that the method
works well for polar III-V semiconductors with zincblende structure:
the calculated cohesive energies are 92$\pm$3\% of the experimental
values using an uncontracted  ($4s4p2d1f$) valence basis set.  Work
is underway in our laboratory to apply the method to other
ground-state properties of the III-V compounds.

\section{Acknowledgments}

We are grateful to Prof.\ H.-J.\ Werner, Stuttgart, and to
Prof.\ R.\ Dovesi, Torino, for providing their programs {\sc Molpro}
and {\sc Crystal} respectively. We also thank Dr.\ T.\ Leininger,
Stuttgart,  and Dr.\ M.\ Dolg, Dresden, for providing the Ga and In
small-core pseudopotentials prior to publication.


\begin{table} \caption{\label{basis} Crystal-optimized basis sets for
the III-V semiconductors.  Of the $s$ and $p$ functions, only the two
outer uncontracted exponents are listed. The inner contracted ones
are chosen to be the same as in the original atom-optimized basis
sets \protect{\cite{bergner93,13vepp}}. }
\begin{tabular}{c||c||cc|cc|c} &&$s$&$s$&$p$&$p$&$d$\\ \hline\hline
BN&B\footnotemark[1]&0.53&0.23&0.53&0.32&0.80\\ \cline{2-7}
&N&0.706251&0.216399&0.808564&0.263&0.82\\ \hline
BP&B\footnotemark[1]&0.53&0.32&0.53&0.375&0.80\\ \cline{2-7}
&P&0.331929&0.120819&0.41&0.16&0.80\\ \hline
BAs&B\footnotemark[1]&0.53&0.17&0.38&0.20&0.34\\ \cline{2-7}
&As&0.294449&0.111896&0.254421&0.15&0.32\\ \hline
AlP&Al&0.170027&0.12&0.20&0.14&0.29\\ \cline{2-7}
&P&0.331929&0.120819&0.227759&0.15&0.48\\ \hline
AlAs&Al&0.170027&0.12&0.20&0.15&0.29\\ \cline{2-7}
&As&0.294449&0.12&0.254421&0.13&0.32\\ \hline
AlSb&Al&0.170027&0.105&0.212&0.143&0.29\\ \cline{2-7}
&Sb&0.260956&0.100066&0.242631&0.083416&0.29\\ \hline
GaP&Ga&0.294700&0.14&0.396817&0.15&0.502918\\ \cline{2-7}
&P&0.331929&0.120819&0.227759&0.15&0.48\\ \hline
GaAs&Ga&0.294700&0.15&0.396817&0.138&0.502918\\ \cline{2-7}
&As&0.294449&0.111896&0.254421&0.107&0.32\\ \hline
GaSb&Ga&0.294700&0.15&0.396817&0.16&0.502918\\ \cline{2-7}
&Sb&0.260956&0.100066&0.242631&0.083416&0.29\\ \hline
InP&In&0.176720&0.082360&0.200692&0.10&0.263986\\ \cline{2-7}
&P&0.331929&0.120819&0.227759&0.13&0.48\\ \hline
InAs&In&0.176720&0.082360&0.200692&0.080578&0.263986\\ \cline{2-7}
&As&0.294449&0.111896&0.254421&0.107&0.32\\ \hline
InSb&In&0.176720&0.082360&0.200692&0.080578&0.263986\\ \cline{2-7}
&Sb&0.260956&0.100066&0.242631&0.083416&0.29\\ \hline
\multicolumn{7}{l}{\footnotesize \footnotemark[1] The very diffuse
$d$-projector of the boron pseudopotential was neglected.}
\end{tabular} \end{table}

\begin{table} \caption{\label{polbasis}Polarization functions used in
the basis sets for the CCSD calculations.} \begin{tabular}{c||c||ccc}
&Basis A, 1d& \multicolumn{3}{c}{Basis B, 2d1f}\\ \hline\hline
B&0.34&0.34& 0.09& 0.49\\ \hline Al&0.29&0.29& 0.10& 0.35\\ \hline
Ga&0.23&0.23& 0.10 & 0.36\\ \hline In&0.12&0.12& 0.04& 0.32\\ \hline
N&0.82& 0.82 &0.23 & 1.09\\ \hline P& 0.48&0.48& 0.16 & 0.56\\ \hline
As& 0.32&0.32& 0.12 & 0.47\\ \hline Sb & 0.19&0.19& 0.07  &0.40
\end{tabular} \end{table}

\begin{table} \caption{\label{inc}Correlation-energy increments for
GaAs (in a.u.), determined at the CCSD level using basis set A. For
the numbering of the clusters and bonds involved, see Fig.
\protect{\ref{fig}} . } \begin{tabular}{l|l|r|l} &Source
cluster/&Increment&Weight factor\\ &bond orbitals&&for the solid\\
\hline $ \epsilon_i$&1/1& -0.019935&\\ &2/1&-0.018800&\\
&7/1&-0.018780&\\ &8/1&-0.018612&4\\ \hline $\Delta
\epsilon_{ij}$&2/1,2&-0.011748&\\ &7/1,2&-0.011445&\\
&8/1,2&-0.010962&6\\ &2/1,3&-0.003155&\\ &7/1,3&-0.003943&\\
&8/1,3&-0.004849&6\\ &8/2,3&-0.000843&12\\ &8/2,5&-0.000568&24\\
&3/1,4&-0.000205&6\\ &3/2,5&-0.000073&6\\ &4/1,4&-0.000158&24\\
&5/2,5&-0.000048&24\\ &4/2,5&-0.000085&12\\ &5/1,4&-0.000109&12\\
&6/1,4&-0.000109&12\\ \hline $\Delta \epsilon_{ijk}$&8/1,2,4&
0.001379&4\\ &8/1,3,5&0.000254&4\\ &8/1,2,3&0.000022&12\\ &8/1,2,5&
0.000124&24 \end{tabular} \end{table}

\begin{table} \caption{\label{result}Cohesive energies per unit cell
(in a.u.), at different theoretical levels (cf.\ text). Deviations
from experimental values (in percent) are given in parentheses. For
comparison, literature  data from DFT calculations are also
reported.  } \begin{tabular}{c||cc|cc|cc|cc|cc|c}
&\multicolumn{2}{c|}{HF}&\multicolumn{2}{c|}{HF+corr}&\multicolumn{2}{c|}
{HF+corr}&
\multicolumn{2}{c|}{LDA\cite{causa94}}&\multicolumn{2}{c|}{GGA\cite{causa94}}
&Expt.\\ &&&\multicolumn{2}{c|}{Basis A}&\multicolumn{2}{c|}{Basis
B}&&&&\\ \hline BN&0.335& (67\%)&0.451&
(90\%)&0.455&(91\%)&0.609&(122\%)&0.505&(101\%)&0.500\\ BP&0.230&
(60\%)&0.325& (85\%)&0.344&(90\%)&&&&&0.382\\
BAs&0.202&&0.303&&0.315&&&&&&\\ AlP&0.198& (64\%)&0.269&
(87\%)&0.292&(94\%)&0.371&(120\%)&0.308&(100\%)&0.309\\ AlAs&0.173&
(59\%)&0.246& (84\%)&0.264&(90\%)&&&&&0.294\\ AlSb&0.146&
(60\%)&0.223& (92\%)&0.232&(95\%)&&&&&0.243\\ GaP&0.147&
(53\%)&0.221& (81\%)&0.246&(91\%)&&&&&0.271\\ GaAs&0.130&
(53\%)&0.207& (84\%)&0.228&(93\%)&0.285&(116\%)&0.229&(93\%)&0.246\\
GaSb&0.109& (49\%)&0.183& (82\%)&0.198&(89\%)&&&&&0.223\\ InP&0.142&
(57\%)&0.205& (82\%)&0.234&(94\%)&&&&&0.250\\ InAs&0.129&
(54\%)&0.194& (81\%)&0.219&(92\%)&&&&&0.239\\ InSb&0.117&
(55\%)&0.182& (85\%)&0.201&(94\%)&&&&&0.213 \end{tabular} \end{table}

\begin{figure} 
\caption{\label{fig}X$_n$Y$_n$H$_m$-clusters  treated
at the  CCSD level. Big numbers designate clusters, small numbers the
bonds in each cluster. H-atoms are not drawn.}
\end{figure}

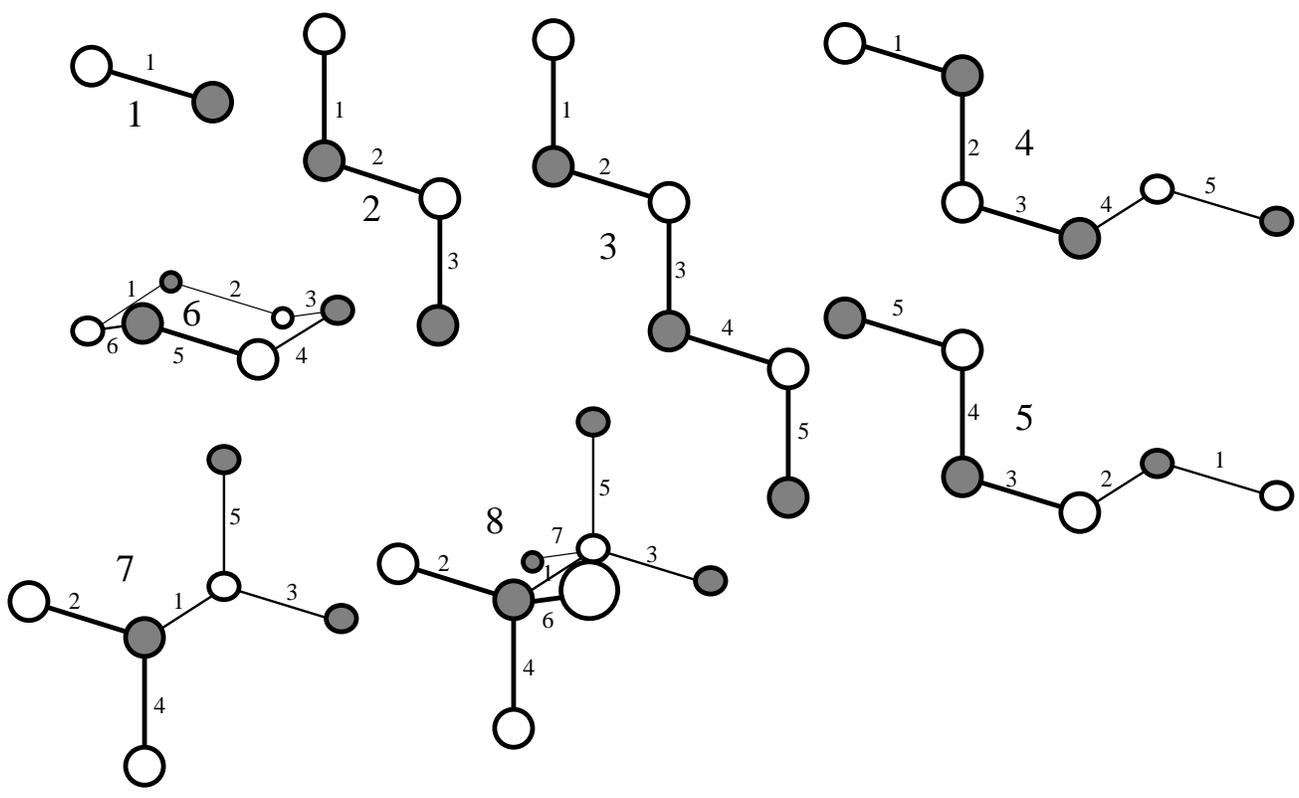
\end{document}